\documentstyle[12pt]{article}
\def\ai{\'{\i}}

\def\-eu{e^{-2U}}
\def\om{\omega}
\def\2om{2\omega +3}
\def\ce{{\cal E}}
\def\8{8\pi G}
\def\pp{\varphi}
\def\4{4\pi G}
\def\16{16\pi G}

\def\be{\begin{equation}}
\def\ee{\end{equation}}
\def\te{{\tilde E}_0}
\def\tm{\tilde M}

\begin{document}

\baselineskip.33in

\centerline{\large{\bf WIGGLY STRINGS IN LINEARIZED BRANS-DICKE GRAVITY}}

\bigskip

\centerline{ Andr\'es Arazi\footnote{Electronic mail: arazi@tandar.cnea.gov.ar} and Claudio Simeone\footnote{Electronic mail: simeone@tandar.cnea.gov.ar}}

\medskip
\centerline{\it Departamento de F\ai sica, Comisi\'on Nacional de Energ\ai a At\'omica}

\centerline{\it Av. del Libertador 8250, 1429 Buenos Aires, Argentina}

\vskip1cm

\noindent ABSTRACT

\bigskip

\noindent The metric around a wiggly cosmic string is calculated in the linear approximation of Brans-Dicke theory of gravitation. The equations of motion for relativistic and non-relativistic particles in this metric  are obtained. Light propagation is also studied and it is shown that photon trajectories can be bounded.

\vskip1cm


\bigskip

\newpage 

\noindent{\bf 1. Introduction}

\medskip

\noindent In the framework of present unified theories a scalar field should exist besides the metric of the spacetime. Scalar-tensor theories of gravitation would be important when studying the early universe, where it is supposed the coupling of the matter to the scalar field could be nonnegligible. Topological defects as cosmic strings are produced in phase transitions in the early universe, so that it seems natural to study their gravitational effects in  a scalar-tensor theory as that of Brans and Dicke [1,2,3,4,5].

In particular, local $U(1)$ cosmic strings [6,7] can have small structure in the form of wiggles and kinks. Far from a string these small structures cannot be resolved, but they modify the average energy per unit length as well as the tension along the string: the energy is larger, while the tension is lower than that of a straight string. An important consequence of this is that the metric (obtained from general relativity)  around a wiggly string is no more flat, but it has $g_{00}\neq 1$ [8,9], so that, differing from straight local strings [10] it can interact with non relativistic matter.   In part for this reason, in recent years wiggly strings have received considerable attention due to the possible role they could play in cosmology, in particular in early structure formation in the universe [8,9]. 

 In the present work  the metric around a wiggly cosmic string is obtained  in the linearized Brans-Dicke theory. It is shown that an appropriate coordinate choice leads to a metric which (up to first order) in a plane normal to the string is conformal to a flat metric with a deficit angle which does not depend on the Brans-Dicke constant $\om$. However,  nearby particles in such a plane develop a  relative acceleration. Also, the motion of relativistic  particles  is studied in the Hamilton-Jacobi formalism. A difference between  wiggly and straight strings is found by analyzing the propagation of light out of a plane normal to the string.

\bigskip
\newpage 

\noindent{\bf 2. The metric}

\medskip

\noindent In Brans-Dicke theory matter and nongravitational fields generate a long-range scalar field $\phi$, which, together with them, acts as a source of gravitational field. The field $\phi$ is a solution of the equation 
\be
\phi_{;\sigma}^{\ ;\sigma}={1\over\sqrt{-g}}{\partial\over\partial x^\sigma}\left(\sqrt{-g}\ g^{\sigma\tau}{\partial\phi\over\partial x^\tau}\right)={8\pi T\over\2om}
\ee
where $T=\delta_\mu^{\ \nu} T_\nu^{\ \mu}$ and $\om$ is a dimensionless constant; the metric equations replacing those of general relativity are 
\be
R_{\mu\nu}-{1\over 2}g_{\mu\nu}R=8\pi {T_{\mu\nu}\over\phi}+{\om\over\phi^2}\phi_{,\mu}\phi_{,\nu}-{\om\over 2\phi^2} g_{\mu\nu}\phi_{,\alpha}\phi^{,\alpha}+{1\over\phi}\phi_{,\mu;\nu}-{1\over\phi}g_{\mu\nu} \phi_{;\sigma}^{\ ;\sigma}.
\ee
In the linear approximation we write $ g_{\mu\nu}=\eta_{\mu\nu}+ h_{\mu\nu}$  
 with $\eta_{\mu\nu}=diag(1,-1,-1,-1)$ in Cartesian coordinates and $h_{\mu\nu}\ll 1$; the Brans-Dicke field is expanded up to  first order as  $\phi\approx\phi_0+\xi= G^{-1}+\xi,$
where $G$ is the gravitational constant. The equations for the metric are then
\be R_{\mu\nu} ^{(1)}=\8 \left(T_{\mu\nu}- {\om+1\over\2om}\eta_{\mu\nu} T\right)+G\xi_{,\mu,\nu}\ee
and the equation for the  field $\phi$ is given by the Minkowskian D'Alembertian \be \Box\phi=\Box\xi={8\pi T\over\2om}.\ee
While for  linearized general relativity a usual gauge choice is $(h_\mu^\nu-\delta_\mu^\nu\ h)_{,\mu}=0$, to solve the linearized Brans-Dicke equations  it is more appropriate   the Brans-Dicke gauge [2]
\be (h_\mu^\nu-\delta_\mu^\nu\ h)_{,\mu}=G\xi_{,\nu}.\ee
 With this gauge choice the equations for a static perturbation $h_{\mu\nu}$ take the simple form
\be\nabla^2 h_{\mu\nu} =\16 \left( T_{\mu\nu}-{\om+1\over\2om}\eta_{\mu\nu}T\right).\ee

Consider a wiggly string along the $z$ axis. Averaging over a distance and a time much greater than the typical wavelength and oscillating period of  the wiggles the energy-momentum tensor can be written in the form [8,9]
\be T_\mu^{\ \nu}=diag ({\ce} ,0,0,-p) \delta(x)\delta(y),\ee
where  
$\ce=\mu+a,\ \  p=-\mu+b,$
with $\mu$ the linear mass density of the unperturbed string and $-p={\cal T}>0$ the tension;  the variations of energy and tension have been estimated [9] as $a\sim 0.4\mu,$   $b\sim 0.3\mu.$
 We must then solve the equations
$$\nabla^2 h_{00}=16\pi G\left({\ce(\om+2)+p(\om+1)\over\2om}\right)\delta(x)\delta(y),$$
$$\nabla^2 h_{11}=\nabla^2 h_{22}=16\pi G(\ce-p)\left({\om+1\over\2om}\right)\delta(x)\delta(y),$$
\be\nabla^2 h_{33}=16\pi G\left({\ce(\om+1)+p(\om+2)\over\2om}\right)\delta(x)\delta(y).\ee  In cylindrical coordinates the resulting metric reads
\begin{eqnarray}ds^2 & = &   \left[1+8G\left({\ce(\om+2)+p(\om+1)\over\2om}\right)\ln(\rho/\rho_0)\right]\, dt^2\nonumber\\ &  &   \ \ \ \ \ \ \ \ \ -\left[1-8G(\ce-p)\left({\om+1\over\2om}\right)\ln(\rho/\rho_0)\right]\, (d\rho^2+\rho^2 d\theta^2)\nonumber\\ & & \ \ \ \ \ \ \ \ \ \ \ \ \  \ \ \ \  -\left[1-8G\left({\ce(\om+1)+p(\om+2)\over\2om}\right)\ln(\rho/\rho_0)\right]\,  dz^2
\end{eqnarray}
with $\rho_0$ a constant of integration (note the analogy with the logarithmic  Newtonian potential for a point mass in  $2+1$ dimensions). In the general relativity limit $\om\ \to\ \infty$ the metric of Reference 7 is recovered.

The Brans-Dicke field, given by the time-independent solution of equation (4) with $T=(\ce -p)\delta (x) \delta (y)$, is 
\be\phi = G^{-1} + {4(\ce -p)\over\2om}\ln(\rho/\rho_0).\ee 
An effective gravitational coupling ``constant'' $G(\rho)$ defined as $\phi^{-1}$  becomes weaker as we go far from the string.
 
To get a better insight of the metric (9) we shall evaluate the Riemann tensor $R^\sigma_{\ \mu\rho\nu}=g^{\lambda\sigma}R_{\lambda\mu\rho\nu}$. 
 The only nonzero components of $R_{\lambda\mu\rho\nu}$ are
$$R_{0101}=-R_{0202}=4G \left({\ce(\om+2)+p(\om+1)\over \2om}\right)\left({1-2\sin^2\theta\over \rho^2}\right)$$
$$R_{1313}=-R_{2323}=
4G\left({\ce(\om+1)+p(\om+2)\over \2om}\right)\left({1-2\sin^2\theta\over \rho^2}\right)$$
$$ R_{0120}=-8G\left({\ce(\om+2)+p(\om+1)\over \2om}\right){\cos\theta \sin\theta\over \rho^2}$$ 
\be R_{1212}=-8\pi G(\ce-p)\left({\om+1\over\2om}\right)\delta(x)\delta(y),
\ee
where $\theta$ is  measured from the $x$ axis. In a plane normal to the string the space metric is flat except at the origin,    in the sense that  the change in the components of a vector $A_{\mu}$ 
in a parallel transport at a fixed time around a closed loop in  a  $xy$ plane  would be zero if the string is not surrounded by the loop. However, the spacetime  metric, even in a $xy$ plane, is not flat, and nearby particles will develop a nonvanishing relative acceleration. 
 For example, for two particles  separated by a distance $X$, and which are both at the same distance from the string moving with the same initial  speed along the $y$ axis, the relative covariant acceleration    is (up to the first order)
\begin{eqnarray*}{D^2 X \over Ds^2} & = & -R_{1010} X \left({dx^0\over ds}\right)^2\\
& = & -\left({4G\over\2om }\right){[\ce(\om+2)+p(\om+1)]\over \rho^2} X \left({dx^0\over ds}\right)^2.
\end{eqnarray*}
As it could be expected, for $p=-\ce$ (straight string) and in the general relativity  limit $\om\ \to\ \infty$  we obtain $D^2 X / Ds^2= 0.$

 If we write 
$ ds^2=g_{00}\left( dt^2+\sum_{i=1} ^3 g^{00}g_{ii} (dx^i)^2\right)$
and redefine the radial coordinate by
\be (1-8G\ce \ln(\rho/\rho_0))\rho^2= (1-8G\ce )r^2,\ \ \ \ \ \ 
 (1-8G\ce \ln(\rho/\rho_0))d\rho^2\approx dr^2,\ee
 neglecting second order terms we can write
\begin{eqnarray}
ds^2 & = & \left[ 1+ 8G\left({\ce(\om+2)+p(\om+1)\over \2om}\right) \ln r\right]\times\nonumber\\ & &  \ \ \ \ \ \ \ \ \ \times  \left(dt^2-dr^2-(1-8G\ce)r^2 d\theta^2    -[1-8G(\ce+p) \ln r]  dz^2\right) .
\end{eqnarray}
We see that  in a plane perpendicular to the $z$ axis the metric is conformal to one with a deficit angle 
$\Delta = 8\pi G\ce .$
 Note that $\Delta$ does not depend on the Brans-Dicke constant $\om$. A new angular variable $\varphi$  such that $d\pp=(1-4G\ce)d\theta$ with $0\leq\varphi\leq 2\pi(1-4G\ce)$ can be defined; in terms of $\pp$ the metric reads 
\begin{eqnarray}
ds^2 & = & \left[ 1+ 8G \left({\ce(\om+2)+p(\om+1)\over \2om}\right) \ln r\right] \times\nonumber\\     & & \ \ \ \ \ \ \ \ \times\left(dt^2-dr^2-r^2 d\pp^2    -[1-8G(\ce+p) \ln r]  dz^2\right).
\end{eqnarray}
In the limit $\om\ \to\ \infty$ and in the case $p=-\mu=-\ce$ the well known Vilenkin's conical metric [10] is obtained. 

\bigskip


\noindent{\bf 3. Trajectories of particles and light}

\medskip

\noindent We shall consider the motion of particles in the metric (14). Because $g_{00}\neq 1$ the wiggly string interacts with particles at rest or with non relativistic speed. The proper acceleration of a non relativistic particle in this metric  is given by
\be {d^2 r\over ds^2}= -{1\over 2} {\partial h_{00}\over \partial r}= -{4G\over r}\left({\ce(\om+2)+p(\om+1)\over\2om}\right)\ee
so that it is attracted by the string.
According to  Vachaspati's [9] estimates for $\ce$ and $p$, 
for $\om\sim 500$ we obtain a  slight correction of about $ 0.3 \% $ above the result  from general relativity.  

The trajectory of a relativistic massive particle   can be obtained starting  from the Hamilton-Jacobi  equation [11]
\be g^{\mu\nu}{\partial S\over\partial x^\mu}{\partial S\over\partial x^\nu}-m^2=0.\ee
As the metric does not depend on $z$ the (covariant) momentum $p_z$ is conserved;  for the motion in a plane perpendicular to the string we have $p_z=0$.  Because the metric is static and axisymmetric  we propose the action 
$  S=-E_0t+M\,\pp+S_r(r) $
where  $E_0$ and $M$ are constants of  motion, so that the Hamilton-Jacobi equation reduces to an equation for the radial term $S_r$:
\be
E_0^2-{M^2\over r^2}-\left(\partial S_r\over\partial r\right)^2-m^2\left(1+{8G\over\2om}[\ce(\om+2)+p(\om+1)] \ln r\right)=0. \ee
The solution is given by the integral
\be S_r=\int dr\sqrt{E_0^2-m^2\left(1+{8G\over\2om}[\ce(\om+2)+p(\om+1)] \ln r\right) -{{\textstyle M^2}\over {\textstyle r^2}}}\ .\ee
The equation for the angle $\pp$ as a function of $r$ yields from $\partial S/\partial M=constant$. Choosing $constant=0$ we have $\pp=-\partial S_r/\partial M$, and then defining $\te\equiv E_0/m$ and $\tm\equiv M/m$ we obtain 
\be\pp(r)=\int {\tm dr\over r^2\sqrt{\te^2- 1-{{\textstyle 8G}\over{\textstyle \2om}}[\ce(\om+2)+p(\om+1)] \ln r -{{\textstyle \tm^2}\over {\textstyle r^2}}}}\ .\ee
To study the radial motion we can define the ``effective potential''
$$\tilde U^2_{eff}(r)=1+{8G\over\2om}[\ce(\om+2)+p(\om+1)] \ln r +{\tm^2\over r^2}$$
which has a minimum for
$$r={\tm\over 2G^{1/2}}\sqrt{{(\2om) \over \ce(\om+2)+p(\om+1)}}\  .$$
Differing  from that corresponding to the Schwarzchild metric, because of the logarithmic form of the string metric the potential $U^2_{eff}(r)$ goes to $+\infty$  for  both $r\to 0$ and  $r\to \infty$; hence, unbounded trajectories would not exist, and the radial motion is limited by $r_{min}$ and $r_{max}$ which are  the solutions of equation (17) with $p_r=-\partial S_r/\partial r =0$.   We should stress, however, that our results are not valid for too large values of $r$, for which the linear approximation based on the assumption $h_{\mu\nu}\ll 1$ fails.

In the case of  light, because for $dz=0$ the metric (14) is conformally flat, a ray within a $xy$ plane will describe a straight line. However, due to the angle deficit $\Delta=8\pi G\ce$, two parallel rays  passing by opposite sides of the string with the same ``impact parameter'' $b$  intersect each other at a distance $ d\approx b/( 4\pi G\ce )$ beyond the string.

A qualitative difference between a wiggly string and a straight one 
is found in the propagation of light  {\em out of the $xy$ plane} ($k_z\neq 0$). For a straight string ($\ce+p=0$) the metric (14) is conformally flat and hence the propagation is rectilinear [2]. This is not the case for a wiggly string: from the identity $ g^{\mu\nu} k_\mu k_\nu =0$ we obtain
\be
k_0^2-k_r^2-{k_\pp ^2\over r^2}-k_z^2 [1+8G(\ce+p)\ln r]=0
\ee
where $k_0,$ $k_\pp$ and $k_z$ are constants of  motion.  This expression is analogous to equation (17) (with $k_z$ playing the role of the mass $m$) so that for the radial motion the same analysis made for a massive particle can be applied. Hence, for a light ray  whose wave vector has a component along the $z$ axis, the radius of its   trajectory remains between  the zeros of equation (20) with $k_r=0$. 
Note that, as $\om$ does not appear in equation (20), this result is also valid in general relativity.

\vskip1cm

\noindent{\bf 4. Discussion}

\medskip

\noindent In the present work we have shown that if an appropriate gauge choice is made the gravitational field of a wiggly string is easy to obtain in linear Brans-Dicke gravity. We have found that as long as the linear approximation remains valid, the trajectory  of a relativistic particle in a plane normal to the string is always bounded. In the case of light, we have shown that there exists a coordinate choice such that the propagation in a $xy$ plane is along a straight line. However, out of this plane the light propagation is qualitatively different from  that corresponding to a straight string.

It should be emphasized that our results are valid as long as $r$ is small enough to consider $h_{\mu\nu}\ll 1$. In fact, unbounded trajectories are posible if what we considered as a string along the $z$ axis  is  actually  a piece of a closed loop; very far from it the motion of a particle could then be solved by considering a metric of the form 
$$ds^2 =  \left[1-{4G{\cal M}\over r}\left({\om+2\over\2om}\right)\right] dt^2  -\left[1+{4G{\cal M}\over r}\left({\om+1\over\2om}\right)\right] (dr^2 + r^2 d\Omega^2)$$
which is a solution of equation (6) for a point source of mass ${\cal M}\sim \mu L$  ($L$ the length of the  loop).  At intermediate distances the loop shape  must be taken into account.

An interesting point to be stressed is that, analogously as the wiggles do, charge carriers moving along a superconducting string [12] increase the energy per unit length and diminish the tension. If the formation of the string and the appearance of the current correspond to successive phase transitions which occur at considerably different temperatures [13], the  current will be low enough so as to neglect the electromagnetic contribution to the energy-momentum tensor. Hence, the energy-momentum tensor of equation (7) could be considered as a reasonable approximation for a superconducting string and its  gravitational field could be estimated by a metric like (9).  

\bigskip

\noindent{\bf Acknowledgment}
\medskip

\noindent We wish to thank F. D. Mazzitelli for reading the manuscript and  making helpful comments.

\bigskip

\newpage

\noindent{\bf References}

\medskip

\noindent 1. A. A. Sen, N. Banerjee and A. Banerjee, Phys. Rev. {\bf D 56}, 3706 (1997).

\noindent 2. A. Barros and C. Romero, J. Math. Phys. {\bf 36}, 5800 (1995).

\noindent 3. C. W. Misner, K. S. Thorne and J. Wheeler, {\it Gravitation}, W. H. Freeman and company, New York (1997).

\noindent 4. S. Weinberg, {\it Gravitation and Cosmology}, John Wiley and sons, New York (1972).

\noindent 5. A. Arazi and C. Simeone, Gen. Rel. Grav. {\bf 32}, 2259 (2000).

\noindent 6. A. Vilenkin, Phys. Rep. {\bf 121}, 263 (1985).

\noindent 7. A. Vilenkin and E. P. S. Shellard, {\it Cosmic Strings and Other Topological Deffects}, Cambridge University Press, Cambridge (1994).

\noindent 8. T. Vachaspati and A. Vilenkin, Phys. Rev. Lett. {\bf 67}, 1057 (1991).

\noindent 9. T. Vachaspati, Phys. Rev. {\bf D 45}, 3487 (1992).

\noindent 10. A. Vilenkin, Phys. Rev. {\bf D 23}, 852 (1981).

\noindent 11. L. D. Landau and E. M. Lifshitz, {\it The Classical Theory of Fields}, Pergamon Press, Oxford (1975).

\noindent 12. E. Witten, Nucl. Phys. {\bf B 249}, 557 (1985).

\noindent 13. R. Brandenberger, B. Carter, A-C. Davis and M. Trodden, Phys. Rev. {\bf D 54}, 6059 (1996).

\end{document}